%% file: main.tex
\documentclass[conference]{IEEEtran}

\usepackage[dvipsnames]{xcolor}
\usepackage{epsfig}
\usepackage{times}
\usepackage{float}
\usepackage{afterpage}
\usepackage{amsmath}
\usepackage{amstext,cite}
\usepackage{amssymb,bm}
\usepackage{latexsym}
\usepackage{color}
\usepackage{graphicx}
\usepackage{amsmath}
\usepackage{amsthm}
\usepackage{graphicx}
\usepackage[center]{caption}
\usepackage{pstricks}

\usepackage{caption}
\usepackage{subcaption} 
\usepackage{booktabs}
\usepackage{multicol}
\usepackage{lipsum}
\usepackage[T1]{fontenc}
\usepackage{hyperref}
\usepackage{aecompl}
\usepackage{mathrsfs}
\usepackage{enumitem}
\usepackage{sidecap}

\usepackage{tikz}
\usetikzlibrary{shapes.multipart}
\usetikzlibrary{fit}
\usetikzlibrary{shapes.geometric,positioning}
\usetikzlibrary{calc}
\usetikzlibrary{patterns.meta}
\usetikzlibrary{arrows.meta} 

\usepackage{soul}
\usepackage{cancel}

\usepackage{tikz}
\usetikzlibrary{shapes.multipart}
\usetikzlibrary{fit}
\usetikzlibrary{shapes.geometric,positioning}
\usetikzlibrary{calc}
\usetikzlibrary{patterns.meta}
\usepackage{pgfplots}
\usetikzlibrary{plotmarks}
\usetikzlibrary{spy}
\usepackage{tikz-cd}
\usepackage[normalem]{ulem}

\input{macros}

\allowdisplaybreaks

\title{Computer-aided Characterization of Fundamental Limits of Coded Caching with Linear Coding}

\begin{document}

\author{
\IEEEauthorblockN{Niccolò Brembilla$^{\dagger}$, Yinbin Ma$^{\dagger}$, Pietro Belotti$^{*}$, Federico Malucelli$^{*}$, and Daniela Tuninetti$^{\dagger}$}
\IEEEauthorblockA{$^{\dagger}$University of Illinois Chicago, Chicago, Illinois, USA;
                  $^{*}$Politecnico di Milano, Milan, Italy;\\ 
Emails:\{nbrem, yma52, danielat\}@uic.edu, pietro.belotti@polimi.it, malucell@elet.polimi.it}
}

\maketitle

\IEEEpeerreviewmaketitle

\input{sections/abstract}
\input{sections/introduction.tex}
\input{sections/problem}
\input{sections/methodology}

\input{sections/results}

\input{sections/conclusions}

\bibliographystyle{ieeetr}
\bibliography{references}
\newpage 
\appendices

\input{sections/appendix}

\end{document}

%% file: macros.tex
\long\def\comment#1{}

\newcommand{\FF}{{\mathbb F}}

\newcommand{\dv}{{\mathbf d}}

\newcommand{\Ac}{{\mathcal A}}

\newcommand{\Sc}{{\mathcal S}}

\newcommand{\qsf}{{\mathsf q}}

\newcommand{\Bsf}{{\mathsf B}}

\newcommand{\Ksf}{{\mathsf K}}

\newcommand{\Msf}{{\mathsf M}}
\newcommand{\Nsf}{{\mathsf N}}

\newcommand{\Rsf}{{\mathsf R}}

\theoremstyle{definition}
\newtheorem*{rem*}{Remark}

\theoremstyle{plain}
\newtheorem{thm}{Theorem}[section]

\providecommand{\definitionname}{Definition}

%% file: sections/abstract.tex
\begin{abstract}
Inspired by prior work by Tian and by Cao and Xu, this paper presents an efficient computer-aided framework to characterize the fundamental limits of coded caching systems under the constraint of linear coding.
The proposed framework considers non-Shannon-type inequalities which are valid for representable polymatroids (and hence for linear codes), and leverages symmetric structure and problem-specific constraints of coded caching to reduce the complexity of the linear program. 
The derived converse bounds are tighter compared to previous known analytic methods, and prove the optimality of some achievable memory-load tradeoff points under the constraint of linear coding placement and delivery. These results seem to indicate that small, structured demand subsets combined with minimal common information constructions may be sufficient to characterize optimal tradeoffs under linear coding.
\end{abstract}

%% file: sections/introduction.tex
\section{Introduction}

Coded caching, a shared-link caching network, was first introduced by Maddah-Ali and Niesen (MAN) in~\cite{maddah2014fundamental}. It leverages the local caches and coordinates their contents so as to enable 
coded multi-casting, thereby achieving significant load saving compared to conventional (uncoded) methods. 
Yu {\em el al.} in~\cite{yu2017exact} improved on the MAN delivery scheme and showed the exact memory-load tradeoffs under uncoded placement, matching the converse of Wan {\em el al.} in~\cite{wan2020index}.
Yu {\em et al.} in~\cite{yu2018characterizing} proved that the optimal load under uncoded placement is order optimal to within a factor of 2.0048 in general.  

To investigate the fundamental limits beyond~\cite{yu2018characterizing}, Tian  in~\cite{tian2018symmetry} initiated a computer-aided exploration by means of a linear programming (LP) outer bounds over the entropic space (i.e., a convex cone of Shannon-type inequalities).
The computer-aided Information-Theoretic Inequality Prover (ITIP) framework, developed by Yeung {\em et al.} in~\cite{yeung2002framework}, is a software tool designed to assist in proving or disproving information theoretic inequalities.
A ITIP-type framework has been used to generate numerical bounds for network coding~\cite{yeung2012first} and distributed coding storage~\cite{tian2014characterizing}. The ITIP formulates an LP problem with constraints imposed by the specific problem (i.e., coded caching in~\cite{tian2014characterizing}) and the entropic space. 
However, the entropic space cannot fully capture all valid information inequalities, such as Zhang-Yeung inequality~\cite{zhang1998characterization}. Those information inequalities not covered by the entropic space are referred to as {\it non-Shannon-type inequalities}. 
Works such as~\cite{dougherty2007networks,dougherty2015achievable,li2023automated} adopted non-Shannon-type inequalities into ITIP.

Linear rank inequalities are a type of non-Shannon-type inequalities that describe representable polymatroids (including linear vector space) and focuses on inequalities valid for matrix rank~\cite{hammer2000inequalities,dougherty2009linear}.
For coded caching, the introduction of linear rank inequalities enabled the characterization of the optimal tradeoff for the case of three users and three files, under the constraint that all coding operations are linear~\cite{cao2020characterizing}.

The limitation of computer-aided ITIP-type framework is that computation time and memory requirements grow exponentially with the number of variables, thus it is often infeasible to solve even a moderate scale problems.
In the context of coded caching, by leveraging inherent symmetries, it was possible to reduce the complexity of the LP converse bound problem and derive novel results for the case of two users; however, even the full three users and three files remained out of reach~\cite{tian2018symmetry}.

\paragraph*{Contributions}
This paper aims to further reduce the complexity of LP problem in~\cite{tian2014characterizing} by exploiting structure in coded caching with linear coding placement. 
We develop an efficient computer-aided framework with the auxiliary variables created by the common information like in~\cite{cao2020characterizing}, The proposed framework yields numerical bounds, to the best of our knowledge, for the largest coded caching problems given our available computational resources.   
We prove the optimality of the YMA scheme for certain memory sizes in systems with an equal number of users and files ranging from four to six. This result establishes optimality under the constraint of linear coding for both placement and delivery, whereas previous results established optimality only under uncoded placement.

\paragraph*{Paper Outline}
The paper is organized as follows. 
Section~\ref{sec:problem} introduces the problem setting and some known prior results.
Section~\ref{sec:Methodology} explains the proposed methodology for LP reduction.
Section~\ref{sec:Results} presents the numerical results of our proposed framework.
Section~\ref{sec:conclusion} concludes this paper.
Additional results and remarks, as well as the used code, can be found at~\cite{Brembilla2025CodedCachingLP}.

\paragraph*{Notation}
We use the following notation convention throughout this paper.
    Calligraphic symbols denote sets, bold symbols vectors, and 
    Sans-serif symbols denote system parameters.
    $|\cdot|$ denotes either the cardinality of a set or the length of a vector.
    For integers $a \leq b$, we let $[a: b] := \{a, a+1, \ldots, b\}$. The shorthand notation $[n]$ denotes a set of $n$ elements, which can be either $[0: n-1]$ or $[1:n]$; which one is usually clear from the context. %
    For a collection $\{Z_1, \ldots Z_n\}$, given a index set $\Sc \subseteq [n]$, we let $Z_\Sc := \{Z_i: i \in \Sc\}$.

%% file: sections/problem.tex
\section{Problem Formulation and Known Results}
\label{sec:problem}
A $(\Ksf, \Nsf)$ coded caching is defined as follows.
\begin{subequations}
    A server stores a library of $\Nsf$ files, denoted by $W_0,\ldots,W_{\Nsf-1}$, each file has $\Bsf$ i.i.d. uniformly random symbols over $\FF_\qsf$, that is,
    \begin{align}
        H(W_\Sc) := H(W_n: n \in \Sc) = |\Sc| \Bsf, 
        \forall \Sc \subseteq [\Nsf].
        \label{eq:fileconstraints}
    \end{align}
$\Ksf$ users are connected to the server via an error-free shared link. Each user has a local cache that can store no more than $\Msf\Bsf$ symbols, where $\Msf \in [0,\Nsf]$. The cashes are denoted as $Z_0,\ldots,Z_{\Ksf-1}$.
There are two phases in a coded caching scheme.
    {\it Placement Phase:} The server populates the caches as a function of the library, that is,
    \begin{align}
        H(Z_k) \leq \Msf\Bsf, \ H(Z_k \mid W_{[\Nsf]}) = 0, \forall k \in [\Ksf]. 
        \label{eq:cacheconstraints}
    \end{align}   
    {\it Delivery Phase}: Each user demands a single file.  %
    We denote the demand vector as $\dv := [d_0, \ldots, d_{\Ksf-1}]$. The server sends a signal $X_\dv$ with at most $\Rsf\Bsf$ symbols a function of the library, 
    that is,
    \begin{align}
        H(X_\dv) \leq \Rsf\Bsf, \ H(X_\dv \mid W_{[\Nsf]}) = 0,
        \forall \dv \in [\Nsf]^{\Ksf}.
        \label{eq:signalconstraints}
    \end{align}
    Each user must recover its desired file from the delivery signal and its own local cache, that is,
    \begin{align}
        H(W_{d_k} \mid Z_k, X_{\dv} ) = 0, \forall k\in[\Ksf], \ \dv \in [\Nsf]^{\Ksf}. 
        \label{eq:decodingcorrectness}
    \end{align}

\label{eq:cachingconstraints}
\end{subequations}

    {\it Performance metric:} we aim to characterize the {\it load} %
    \begin{align}
        \Rsf^\star(\Msf) = \limsup_{\Bsf \rightarrow \infty} \min_{X, Z_1, \ldots Z_\Ksf} \max_{\dv} \{\Rsf: \text{\eqref{eq:cachingconstraints} is satisfied.}\} \label{eq:worstcaseload}
    \end{align}

Known achievable schemes have the following performance.
\begin{thm}[{YMA scheme~\cite{yu2017exact,yu2018characterizing}}]
        \label{thm:performanceYMA}
    The lower convex envelope of the following points is achievable,
    \begin{align}
        ( \Msf_t, \Rsf_t)^\text{\rm{(YMA)}}  = \biggl(
        \frac{\Nsf t}{\Ksf}, %
        \frac{\binom{\Ksf}{t+1} - \binom{\Ksf -\min(\Ksf,\Nsf)}{t+1}}{\binom{\Ksf}{t}} \biggr), %
        \label{eq:performanceYMA}
    \end{align}
    for every $t \in [0:\Ksf]$.
    Furthermore, the YMA scheme is optimal under the constraint of uncoded placement; and to within a multiplicative factor of two otherwise.
\end{thm}
\begin{thm}[{Gomez scheme~\cite{gomez2018fundamental}}]
        \label{thm:performanceGV}
    When $\Ksf \geq \Nsf$,
    the lower convex envelope of the following points is achievable,
    \begin{align}
        ( \Msf_t, \Rsf_t)^\text{\rm{(GV)}}  = \biggl(
        \frac{\Nsf}{\Ksf t}, %
        \Nsf - \frac{\Nsf(\Nsf+1)}{\Ksf(t+1)} \biggr), %
        \label{eq:performanceGV}
    \end{align}
    for every $t \in [\Ksf]$.
    When $t \geq \Ksf-1$, the Gomez scheme achieves the optimal load.
\end{thm}

We shall use in the following converse bounds from~\cite{yu2018characterizing}, which hold without any restrictions of the placement or delivery phases. 
\begin{thm}[{\cite[Theorem~2]{yu2018characterizing}}]
     \label{thm:converseyutheorem2}
     $\Rsf^\star$ is lower bounded by the lower convex envelope of the following points,
     \begin{align}
         (\Msf, \Rsf) = \left(\frac{\Nsf-\ell+1}{s}, \frac{s-1}{2}+\frac{\ell(\ell-1)}{2s}\right), \notag \\ \forall s \in [\min(\Nsf,\Ksf)], \ell \in [s].
         \label{eq:converseyutheorem2}
     \end{align}
\end{thm}
\begin{thm}[{\cite[Theorem~4]{yu2018characterizing}}]
     \label{thm:converseyutheorem4}
     $\Rsf^\star(\Msf)$ is lower bounded %
     \begin{align}
          \begin{cases}\frac{2 K-n+1}{n+1}-\frac{K(K+1)}{n(n+1)} \cdot \frac{M}{N} & \text { if } \beta+\alpha \frac{K-2 n-1}{2} \leq 0, \\ \frac{2 K-n+1}{n+1}-\frac{2 K(K-n)}{n(n+1)} \cdot \frac{M}{N-\beta} & \text { otherwise, }\end{cases}
          \label{eq:converseyutheorem4}
     \end{align}
     for any $n \in\{\max \{1, \Ksf-\Nsf+1\}, \ldots, \Ksf-1\}$, where $\alpha= \left\lfloor\frac{\Nsf-1}{\Ksf-n}\right\rfloor$ and $\beta=\Nsf-\alpha(\Ksf-n)$.
\end{thm}

Other converse bounds without restrictions on the placement are~\cite{sengupta2017improved,wang2018improved}. To our best knowledge, the combination of Theorem~\ref{thm:converseyutheorem2} and Theorem~\ref{thm:converseyutheorem4}  provides the best converse bound for the general settings, which we focus on in this paper.

In the rest of paper, we improve the computer-aided framework from previous works~\cite{tian2018symmetry,cao2020characterizing}, so we attain some optimal tradeoff memory-load pairs under the constraint of linear coding placement and delivery phases.

%% file: sections/methodology.tex
\section{Methodology for LP reduction}
\label{sec:Methodology}
Following~\cite{tian2018symmetry}, the ITIP framework~\cite{yeung2002framework} can be leveraged as follows to derive converse bounds on coded caching. 
We consider the ${n}:=\Nsf+\Ksf+\Nsf^\Ksf$ discrete random variables in 
$
\mathcal{V} := \{V_1,...,V_n \} = 
W_{[\Nsf]}
\cup Z_{[\Ksf]}
\cup X_{[\Nsf]^\Ksf}
$
We associate the LP-variable $H(V_\mathcal{S})$ to each of the $2^{n} = 2^{\Nsf+\Ksf+\Nsf^\Ksf}$ subsets $\mathcal{S} \subseteq \mathcal{V}$, with the convention $H(V_\emptyset) = 0$.
The elemental Shannon-type inequalities are as follows, for all $\{i,j\} \cup \Ac \subseteq [{n}]$
\begin{subequations}
\begin{align}
    &0\leq H(V_i \mid V_\Ac) 
    = H(V_{\{i\} \cup \Ac}) - H(V_{\Ac}), 
    \\
    &0\leq I(V_i; V_j \mid V_\Ac) 
    = 
    H(V_{\{i\} \cup \Ac}) + H(V_{\{j\} \cup \Ac})
    \notag\\&\quad
    - H(V_{\{i,j\} \cup \Ac})
    - H(V_{\Ac}).
\end{align}
\label{eq:elemental inequalities}
\end{subequations}
In addition to the elemental inequalities in~\eqref{eq:elemental inequalities}, we need to add the problem-specific inequalities in~\eqref{eq:cachingconstraints}.
The LP-objective is the minimization of $\Rsf$ from~\eqref{eq:worstcaseload}.

The complexity of this LP scales at least exponentially in $n$, or equivalently, double exponentially in $\Ksf$. Consequently, to solve the problem within a reasonable time and on reasonable computing platform, we must reduce both the number of LP-variables and the number of LP-constraints. We follow~\cite{tian2018symmetry} to do so, by exploiting the user- and file-symmetry properties. 

In addition, we augment the set of LP constraints by introducing non-Shannon-type (rank) inequalities that are satisfied by random variables corresponding to subspaces of linear vector spaces, such as those generated by linear codes.

\subsection{Exploitation of Symmetries}

Motivated by the goal of developing a more systematic approach to LP simplification, we apply two forms of reduction: (a) \emph{constraint pruning}, (b) \emph{variable merging}, and (c) \emph{demand types}, guided by the user– and file-symmetries~\cite{tian2018symmetry}.
Namely, the caching problem is inherently symmetric with respect to permutations of user and file indices, thus there is no loss of optimality in restricting the search to symmetric coding schemes. %
Consider the file (resp. user) permutation $\hat{\pi}$ (resp. $\bar{\pi}$), we permute the index of file (resp. user) as follows,
\begin{subequations}
\begin{align}
    (0,1,\ldots,\Nsf-1) &\rightarrow\{\hat{\pi}(0),\hat{\pi}(1),\ldots,\hat{\pi}(\Nsf-1)\}, \\
    (0,1,\ldots,\Ksf-1) &\rightarrow\{\bar{\pi}(0),\bar{\pi}(1),\ldots,\bar{\pi}(\Ksf-1)\}.
\end{align}
\label{eq:defofpermutation}
\end{subequations}
The joint entropy are unchanged if their elements are permuted, that is,
$
H(W_\mathcal{A},Z_\mathcal{B}, X_\mathcal{C})
= 
H(
W_{\hat{\pi}[\mathcal{A}]},
Z_{\bar{\pi}[\mathcal{B}]},
X_{\hat{\pi}\circ \bar{\pi} [\mathcal{C}]}
)
$
where $\hat{\pi}\circ \bar{\pi}$ is the permutation of the demand vector indexes as a result of the `composing' the file $\hat{\pi}$ and user $\bar{\pi}$ permutations~\cite{tian2018symmetry}.
With this, we reduce the complexity of LP problem as follows.

\paragraph{Constraint Pruning}  

Each LP-constraint / elemental inequality in~\eqref{eq:elemental inequalities}
involves subsets of the entropy LP-variables.  
To identify and remove constraints that are ``duplicate'' due to user or file symmetries, we generate all admissible user–file permutations and test whether a given inequality can be transformed into one that has already been included; if so, the duplicate inequality is discarded; if not, we add the inequality to the list of LP-constraints.
    Consider for example the 3-user and 3-file (\(3\mathrm{U}3\mathrm{F}\)) setting;
    the constraint 
    $
        H(W_0) \le H(W_0, Z_0)
    $
    is equivalent, under user-file symmetry $\hat{\pi}=\{1,0,2\}$, to
    $
        H(W_1) \le H(W_1, Z_0). 
    $
    Because the two inequalities describe identical dependency relationships between file and cache variables, only the representative with the smallest index, here, $H(W_0) \le H(W_0, Z_0)$, is retained.
    This observation enables us to systematically detect and eliminate many constraints, which in higher-dimensional systems can number in the hundreds of thousands, thereby substantially reducing the load of the LP-solver.

\paragraph{Variable Merging}  

In addition to constraint pruning, further dimensionality reduction is achieved by merging symmetrically equivalent entropy variables.  
  
Here, variables that are symmetric under user or file permutations (e.g., $H(Z_0)$ and $H(Z_1)$ in the above example) represent identical random structures and can therefore be replaced by a single representative LP-variable. 
This reduction directly shrinks the LP’s variable space, which in turn reduces both the number of constraints and the overall computational complexity.

\paragraph{Demand Types}
Rather than evaluating all $\Nsf^\Ksf$ possible  demand configurations, it is sufficient to consider demand types~\cite{tian2018symmetry}, 
defined as a vector of size $\Nsf$ in a decreasing order, $(t_1,\ldots,t_{\Nsf})$ where $t_i \in[0:\Ksf]$ denotes the number of users demanding the $i$-th most demanded file, and $\sum_{i} t_i = \Ksf$. 
For instance, in the \(3\mathrm{U}3\mathrm{F}\) setting, a demand type $(2,1,0)$ indicates that two users request the same file, one user requests another file, and no one requests the third file.

\begin{figure}
\centering
\begin{tikzpicture}[row sep=4em, column sep=4em, every node/.style={font=\small}, every label/.style={font=\small}]
\node (B1) at (0,0) {$\{W_0, Z_1, X_{012}\}$};
\node (C1) at (4.5,0) {$\{W_0, Z_2, X_{120}\}$};

\node (B2) at (0,-2) {$\{W_1, Z_1, X_{102}\}$};
\node (C2) at (4.5,-2) {$\{W_1, Z_2, X_{021}\}$};

\draw[->] (B1) -- node[above] {$\bar{\pi} = \{2,0,1\}$} (C1);
\draw[->] (B1) -- node[left] {$\hat{\pi} = \{1,0,2\}$} (B2);
\draw[->] (C1) -- node[right] {$\hat{\pi} = \{1,0,2\}$} (C2);
\draw[->] (B2) -- node[above] {$\bar{\pi} = \{2,0,1\}$} (C2);
\end{tikzpicture}
\caption{A symmetric example on $\Ksf=\Nsf=3$. Permutation on $X_\star$ refers to~\cite{tian2018symmetry}.}
\label{fig:symexample3u3f}
\end{figure}
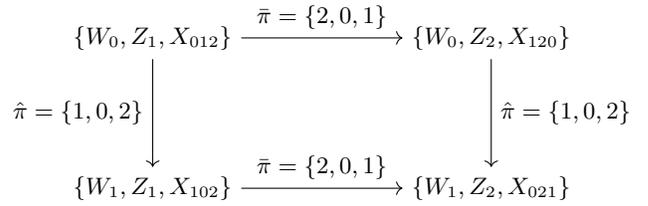

Two subsets of demands are said to be \emph{symmetrically equivalent} if one can be transformed into the other by a combination of user and file permutations.
Under such transformations, the induced LP-variables have identical entropy.
For example, consider again the $3\mathrm{U}3\mathrm{F}$ problem. The demand subsets $\{X_{012}, X_{120}\}$ and $\{X_{102}, X_{021}\}$ are symmetrically equivalent because the file permutation $\hat{\pi} = \{1,0,2\}$ map one subset to the other -- see  Fig.~\ref{fig:symexample3u3f}, which shows that both pairs of subsets are related by the same combination of user and file permutations. In~\cite{tian2018symmetry}, the authors noted that handling all demand types simultaneously for the \(3\mathrm{U}3\mathrm{F}\) problem was infeasible due to memory constraints. Instead, they considered one demand type at a time, solved the corresponding LP, and obtained the overall converse bound as the point-wise maximum of these individual bounds.
Here, by identifying and grouping symmetrically equivalent demand subsets, we can retain only one representative subset from each equivalence class. We then solve one LP per representative subset and combine the resulting rates as in~\cite{tian2018symmetry}. This strategy greatly reduces the number of separate LPs that we need to solve while obtained a `good' aggregated memory-load characterization.

\paragraph{Further Considerations from Decoding Constraints}
Together, the above considerations allow us to make it computationally feasible to handle LP formulations that would otherwise be intractable, thereby extending the range of coded caching configurations that can be explored within reasonable runtime and memory limits. We can further simplify the LP by incorporating the following decoding-constraint consideration.

In general, given $\mathcal{A}\subseteq [\Nsf]$, $\mathcal{B}\subseteq [\Ksf]$ and $\mathcal{C}\subseteq [\Nsf]^\Ksf$, we have
\begin{align}
H(W_\mathcal{A},Z_\mathcal{B}, X_\mathcal{C}) 
&\leq H(W_{[\Nsf]},Z_\mathcal{B}, X_\mathcal{C}) 
\\&= H(W_{[\Nsf]}) + H(Z_\mathcal{B}, X_\mathcal{C} \mid W_{[\Nsf]})
\\&= \Nsf \Bsf + 0,
\end{align}
with equality  if
\begin{equation}
    \cup_{u\in\mathcal{B}, \dv\in \mathcal{C}} \{[\dv]_u\} \cup \mathcal{A} = [\Nsf].
\label{eq:entropySisN}
\end{equation}
That is, \eqref{eq:entropySisN} captures the fact that any collection of variables sufficient to recover the entire file library contains exactly the total information of all $\Nsf$ files. 
If that is the case, then we can simply set 
$H(W_\mathcal{A},Z_\mathcal{B}, X_\mathcal{C}) =\Nsf \Bsf$, thereby reducing the number of degrees of freedom and improving solver efficiency without altering the validity of the bound.

\subsection{Integration of Common Information (CI)}

When only Shannon-type inequalities are imposed, the resulting LP yields a valid converse bound that holds under \emph{any} (potentially nonlinear) placement and delivery scheme.  
However, using Shannon-type inequalities alone is not sufficient to characterize all information inequalities, particularly in problems such as network coding and index coding~\cite{dougherty2007networks}.  
In the context of coded caching, this limitation was made explicit in the 3U3F setting, where \cite{cao2020characterizing} demonstrated that the optimal load under \emph{linear} coded placement (and delivery) can be exactly characterized by augmenting the LP with additional random variables, referred to as \emph{Common Information (CI)} variables, that capture shared linear structure among the involved random variables. 
These additional CI constraints effectively restrict the converse to hold under the assumption that both the placement and delivery phases employ \emph{linear} coding~\cite{dougherty2009linear,cao2020characterizing}.

Motivated by this approach, we extend the LP formulation by introducing analogous CI variables for larger system configurations.  
A random variable \(C\) is said to be a \emph{common information} of random variables \(A\) and \(B\) if it is a deterministic function of both and satisfies
\begin{align}
H(C|A) = 0, \
H(C|B) = 0, \
H(C) = I(A;B).
\end{align}
For simplicity, we write such $C$ as \(\{A;B\}\) when the context is clear. 
For coded caching with linear coding, such CIs always exist because the associated random variables correspond to subspaces of a finite-dimensional vector space, and their intersections naturally define deterministic shared components.

By introducing auxiliary CIs that represent the common information among selected groups of files, cache contents, and delivery messages, the LP formulation is enlarged to include both these additional variables and the corresponding constraints.  
This extended variable set enables the LP to capture linear rank inequalities that are non-Shannon-type with respect to the original entropy variables.  
Incorporating CIs is therefore essential to reproduce the known optimal memory-load points in the $3\mathrm{U}3\mathrm{F}$ setting reported in~\cite{cao2020characterizing}, and, as our experiments suggest, the same modeling principle can be effectively applied to larger configurations to obtain comparably tight bounds under linear coding assumptions.

%% file: sections/results.tex
\section{Numerical Results}
\label{sec:Results}

In this section we present the numerical results of improved LP problem by complexity reduction approaches in Section~\ref{sec:Methodology}. 
Our analysis targets the memory regime
\begin{equation}
\Msf \in \left[\frac{1}{\Nsf-1}, \left\lfloor \frac{\max(\Nsf,\Ksf)}{2} \right\rfloor \right],
\end{equation}
which remains the least characterized region of the memory–rate tradeoff, particularly for
\begin{equation}
\Ksf \ge 4 \quad \text{or} \quad \Nsf \le \frac{\Ksf(\Ksf+1)}{2}.
\end{equation}
We report detailed results for symmetric instances with \(\Nsf=\Ksf \in\{4,5,6\}\). 
These configurations represent, to the best of our knowledge, the first work to %
derived converse bounds for systems of this scale under linear coding assumptions.  
The resulting bounds therefore reflect the tightest rates obtainable under these computational constraints, while maintaining methodological consistency across all tested instances.

All experiments were executed on a single node of Delta at NCSA, a high-performance computing cluster using the Gurobi~12.0.1 solver. The node features two AMD~EPYC~7763 (“Milan”) processors (128~cores total at 2.45~GHz) and 256~GB of DDR4~RAM.  
The system operates under a NUMA configuration with four domains per socket (\texttt{NPS=4}), improving memory locality and bandwidth utilization during large-scale LP optimization. Gurobi was configured to exploit intra-domain parallelism, minimizing cross-domain memory access and ensuring stable performance across problem instances.

In the following, the figures report the difference between the achievable load from the best performing scheme between Theorems~\ref{thm:performanceYMA} and ~\ref{thm:performanceGV}, and various converse bounds, including both the numerically obtained bound and those from Theorems~\ref{thm:converseyutheorem2} and ~\ref{thm:converseyutheorem4}.

\subsection{4U4F: Optimality at $M=1$}
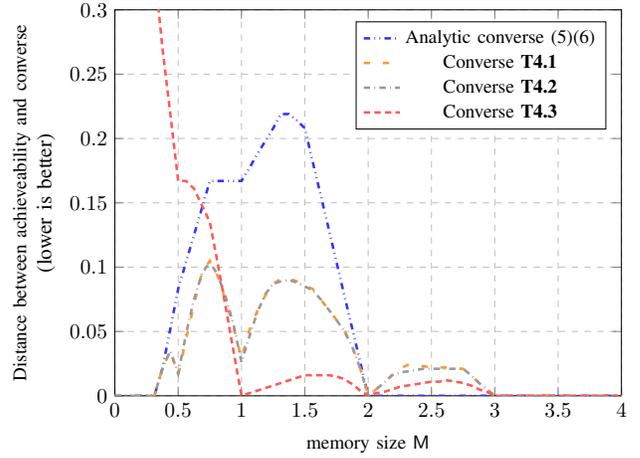
\begin{figure}%
  \centering
  \scalebox{0.8}{
  \begin{tikzpicture}
        \begin{axis}[
            xmin=-0.0, xmax=4.0,
            ymin=-0.0, ymax=0.3,
            legend entries={ 
                \small Analytic converse~\eqref{eq:converseyutheorem2}\eqref{eq:converseyutheorem4},
                \small Converse~\ref{p:t4t1},
                \small Converse~\ref{p:t4t2},
                \small Converse~\ref{p:t4t3}
            },
            xtick={0,0.5,...,4},
            ytick={0,0.05,...,0.3},
            y tick label style = {
                /pgf/number format/fixed,
                /pgf/number format/precision=2
            },
            height=8cm,
            width=10cm,
            grid=major,
            grid style=dashed,
            legend pos=north east,
            ylabel near ticks,
            xlabel={\small memory size $\Msf$},
            ylabel style={align=center},
            ylabel={\small Distance between achieveability and converse \\ (lower is better)},
            scaled ticks=false
        ]
        \addplot[color=blue!80, dash dot dot, very thick] table [x=x, y=y1] {data/4u4f.data};
        \addplot[color=orange!80, loosely dashed, very thick] table [x=x, y=y2] {data/4u4f.data};
        \addplot[color=gray!80, dashdotted, very thick] table [x=x, y=y3] {data/4u4f.data};
        \addplot[color=red!65, densely dashed, very thick] table [x=x, y=y4] {data/4u4f.data};
        \end{axis}
    \end{tikzpicture}
    }
  \caption{4U4F case: achievability is derived from~\eqref{eq:performanceYMA}-\eqref{eq:performanceGV}. \ref{p:t4t3} proves that YMA scheme is optimal at $\Msf=1$.} 
  \label{fig:4u4f}
\end{figure}
For the 4-user, 4-file case, several tradeoffs were evaluated in Fig.~\ref{fig:4u4f}. The most relevant configurations are summarized below.
\begin{enumerate}[label=\textbf{T4.\arabic*}]
\item  \label{p:t4t1} Demands
\(\{X_{0123}, X_{0132}, X_{0213}, X_{1023}\}\) and auxiliary CI  
 \(\{W_0, W_1; X_{0123}\}\), \(\{ X_{0123},Z_1;W_2,W_3,Z_3\}\).
\item \label{p:t4t2} Demands
\(\{X_{0123}, X_{0132}, X_{0213}, X_{1023}\}\) and auxiliary CI \(\{W_0, W_1;X_{0123}\}\), \(\{W_1, Z_2; W_0, X_{0123}\}\).
\item \label{p:t4t3} Demands 
\(\{X_{0012}, X_{0131}, X_{0211}\}\) and auxiliary CI 
\(\{W_0, W_1, Z_1;W_2, X_{0131}, Z_2\}\), \(\{W_0, W_3;X_{0131}, Z_1\}\).
\end{enumerate}

Tradeoffs~\ref{p:t4t1} and~\ref{p:t4t2}, whose demand type is $(1,1,1,1)$ (referred to as the all-file-demanded type), dominate the very-low-memory regime ($\Msf \le 0.875$), while Tradeoff~\ref{p:t4t3} %
governs the intermediate range $\Msf \in [0.875, 2]$. 

Notably, Tradeoff~\ref{p:t4t3} matches the YMA scheme (Theorem~\ref{thm:performanceYMA})  at $\Msf = 1$, achieving a zero gap and thus demonstrating its optimality \emph{under linear coding for both placement and delivery phases}. 
{\bf
This improves upon the previous result showing that YMA is optimal only under uncoded placement. It also implies that, if the Shannon-type inequality–based converse bound is indeed tight, any further improvement would require exploring non-linear coding schemes, which are poorly understood at present.
}

Importantly, this structure can be realized with a minimal construction: a single representative demand of type $(2,1,1,0)$, without any CI, suffices to reach the matching point at $\Msf = 1$. 

\subsection{5U5F: Optimality at $M=1$ }
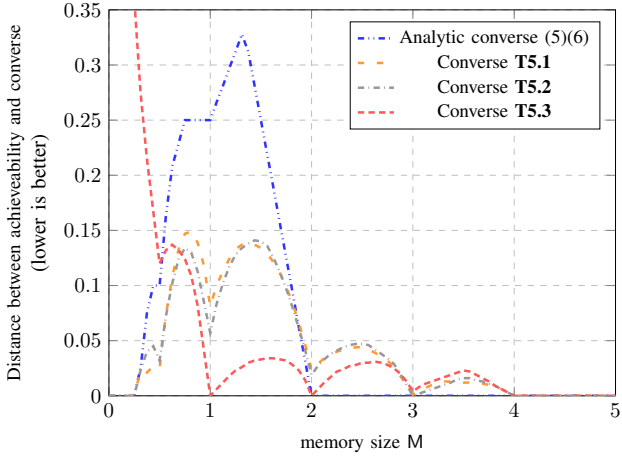
\begin{figure}%
  \centering
  \scalebox{0.8}{
  \begin{tikzpicture}
        \begin{axis}[
            xmin=-0.0, xmax=5.0,
            ymin=-0.0, ymax=0.35,
            legend entries={ 
                \small Analytic converse~\eqref{eq:converseyutheorem2}\eqref{eq:converseyutheorem4},
                \small Converse~\ref{p:t5t1},
                \small Converse~\ref{p:t5t2},
                \small Converse~\ref{p:t5t3}
            },
            xtick={0,1,...,5},
            ytick={0,0.05,...,0.35},
            y tick label style = {
                /pgf/number format/fixed,
                /pgf/number format/precision=2
            },
            height=8cm,
            width=10cm,
            grid=major,
            grid style=dashed,
            legend pos=north east,
            ylabel near ticks,
            xlabel={\small memory size $\Msf$},
            ylabel style={align=center},
            ylabel={\small Distance between achieveability and converse \\ (lower is better)},
            scaled ticks=false
        ]
        \addplot[color=blue!80, dash dot dot, very thick] table [x=x, y=y1] {data/5u5f.data};
        \addplot[color=orange!80, loosely dashed, very thick] table [x=x, y=y2] {data/5u5f.data};
        \addplot[color=gray!80, dashdotted, very thick] table [x=x, y=y3] {data/5u5f.data};
        \addplot[color=red!65, densely dashed, very thick] table [x=x, y=y4] {data/5u5f.data};
        \end{axis}
    \end{tikzpicture}
    }
  \caption{5U5F case: achievability is derived from~\eqref{eq:performanceYMA}-\eqref{eq:performanceGV}. \ref{p:t5t3} proves that YMA scheme is optimal at $\Msf=1$.}
  \label{fig:5u5f}
\end{figure}
For the 5-user, 5-file case, three dominant tradeoffs were identified in Fig.~\ref{fig:5u5f}.

\begin{enumerate}[label=\textbf{T5.\arabic*}]
\item \label{p:t5t1} Demands
\(\{X_{01234}, X_{32401}, X_{42130}, X_{13240}, X_{23401}\), \(X_{43210}\}\) and auxiliary CI  
\(\{W_0, W_1;  W_2, X_{01234}\}\), \(\{W_0, W_1; W_2, W_3, X_{01234}\}\).
\item \label{p:t5t2} Demands
\(\{X_{01234}, X_{01342}, X_{02314}, X_{04123}, X_{04312}\}\) and auxiliary CI \(\{W_0, Z_1;  X_{01234}\}\) and \(\{W_0, W_1;  X_{01234}\}\).
\item \label{p:t5t3} Demands 
\(\{X_{00123}, X_{14003}, X_{01230}\}\) and auxiliary CI 
\(\{W_0, W_1; X_{00123}\}\) and \(\{W_0, Z_0;  X_{00123}, Z_3\}\).
\end{enumerate}

\medskip

Tradeoffs~\ref{p:t5t1} and~\ref{p:t5t2} (all-file-demanded type) dominate the low-memory region, while Tradeoff~\ref{p:t5t3} %
achieves an exact match with the achievable rate of the YMA scheme (Theorem~\ref{thm:performanceYMA}) 
at \(\Msf=1\). This confirms that, at \(\Msf=1\), our converse is tight and therefore the YMA scheme is optimal \emph{under the assumption of linear coding for both placement and delivery}.
Two equivalent minimal constructions yield zero gap at this point:
\begin{itemize}
  \item \textbf{Option~A (3 demands, 1 CI):} three demands of type $(2,1,1,1,0)$ combined with a single CI (as in Tradeoff~\ref{p:t5t3}) match achievability at $\Msf=1$;
  \item \textbf{Option~B (2 demands, 2 CIs):} two demands of the same type with two distinct CI configurations achieve the same optimal point.
\end{itemize}
The combination of ~\ref{p:t5t1}–~\ref{p:t5t3} markedly reduces the gap in the range \(\Msf \in [1,2]\) (to below~0.05), representing, to the best of our knowledge, one of the most refined converse bounds available for the $\Ksf=\Nsf=5$. %

\subsection{6U6F: Optimality at $\Msf=2$ and Near-Optimality at $\Msf=1$}
For the 6-user, 6-file scenario, we focused on optimizing the converse around $\Msf = \{1,2\}$, but those configurations can be used to derive a bound for the whole memory regime. The most relevant tradeoffs are outlined below.
\begin{enumerate}[label=\textbf{T6.\arabic*}]
\item \label{p:t6t1} Demands
\(\{X_{012345}, X_{012453}, X_{021345}, X_{013524}\}\) and auxiliary CI  
\(\{W_0, Z_0; X_{012345}, Z_1\}\), \(\{W_0, Z_1; X_{012345}, Z_2\}\).
\item \label{p:t6t2} Demands
\(\{X_{001234}, X_{010245}, X_{045201}\}\) and auxiliary CI \(\{W_0, W_1;  W_2, X_{001234}\}\).
\item \label{p:t6t3} Demands 
\(\{X_{001234}, X_{012353}, X_{005312}\}\) and auxiliary CI 
\(\{W_0, W_1;  W_2, X_{001234}\}\).
\item \label{p:t6t4} Demands \(\{X_{001234}, X_{001345}, X_{004523}\}\) and auxiliary CI \(\{W_0, W_1;  W_2, X_{001345}\}\).
\end{enumerate}

Due to computational constraints, $(2,1,1,1,1,0)$ configurations were limited to a single CI per tradeoff. Each configuration dominates a distinct memory interval: ~\ref{p:t6t1} (very low $\Msf$), ~\ref{p:t6t3} ($\Msf \in [0.7, 1.4]$), ~\ref{p:t6t2} ($\Msf \in [1.4, 1.875]$), and ~\ref{p:t6t4} ($\Msf \gtrsim 2$). %
The combination of~\ref{p:t6t1}–\ref{p:t6t4} yields, to the best of our knowledge, the tightest converse obtained for $\Ksf=\Nsf=6$ to date.
With only three demands and one CI, Tradeoff~\ref{p:t6t2} achieves an exact match with the YMA achievable rate at $\Msf=2$, thus establishing optimality of the YMA scheme \emph{under the assumption of linear coding for both placement and delivery phases}. At $\Msf=1$, the gap remains small and may potentially be closed through additional CI exploration or expanded demand sampling.

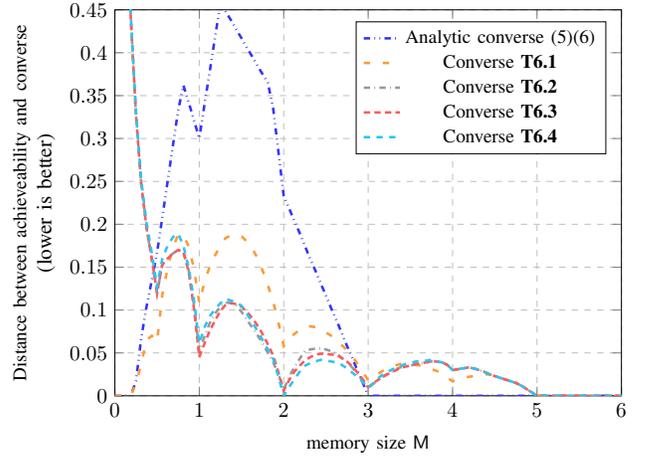
\begin{figure}%
  \centering
  \scalebox{0.8}{
  \begin{tikzpicture}
        \begin{axis}[
            xmin=-0.0, xmax=6.0,
            ymin=-0.0, ymax=0.45,
            legend entries={ 
                \small Analytic converse~\eqref{eq:converseyutheorem2}\eqref{eq:converseyutheorem4},
                \small Converse~\ref{p:t6t1},
                \small Converse~\ref{p:t6t2},
                \small Converse~\ref{p:t6t3},
                \small Converse~\ref{p:t6t4}
            },
            xtick={0,1,...,6},
            ytick={0,0.05,...,0.45},
            y tick label style = {
                /pgf/number format/fixed,
                /pgf/number format/precision=2
            },
            height=8cm,
            width=10cm,
            grid=major,
            grid style=dashed,
            legend pos=north east,
            ylabel near ticks,
            xlabel={\small memory size $\Msf$},
            ylabel style={align=center},
            ylabel={\small Distance between achieveability and converse \\ (lower is better)},
            scaled ticks=false
        ]
        \addplot[color=blue!80, dash dot dot, very thick] table [x=x, y=y1] {data/6u6f.data};
        \addplot[color=orange!80, loosely dashed, very thick] table [x=x, y=y2] {data/6u6f.data};
        \addplot[color=gray!80, dashdotted, very thick] table [x=x, y=y3] {data/6u6f.data};
        \addplot[color=red!65, densely dashed, very thick] table [x=x, y=y4] {data/6u6f.data};
        \addplot[color=cyan!65, dashed, very thick] table [x=x, y=y5] {data/6u6f.data};
        \end{axis}
    \end{tikzpicture}
    }
  \caption{6U6F case: achievability  from~\eqref{eq:performanceYMA}-\eqref{eq:performanceGV}.  Tradeoffs~\ref{p:t6t3} and~\ref{p:t6t4} prove that YMA scheme is optimal at $\Msf=2$, and appear to get close to optimality at $\Msf=1$.}
  \label{fig:6u6f}
\end{figure}

\subsection{Remarks}
Across all evaluated configurations, from $\Ksf=\Nsf=4$ up to $\Ksf=\Nsf=6$, a {\it consistent structural pattern} emerges. Identity-type tradeoffs dominate in the very-low memory regime, while progressively richer non-identity demand types become active as memory increases. This recurrent behavior across distinct settings suggests an underlying organizational principle in the linear-coded caching landscape.

Another notable trend is that {\it relatively compact configurations suffice}, that is, often configurations involving only two-demand instances with a single CI variable are sufficient to attain near-optimal or even exact converse points (e.g., $\Msf = 1$ for $\Ksf\in\{4,5\}$ and $\Msf=2$ for $\Ksf=6$). The same qualitative pattern is visible in higher-dimensional experiments (\emph{cf.}~repository), though these results are omitted here due to space constraints.

The execusion time per LP instance ranged from a few minutes for $\Nsf{=}\Ksf{=}4$ to several hours for $\Nsf{=}\Ksf{=}6$, depending on the degree of symmetry reduction and the number of auxiliary CI variables. These measurements provide a representative indication of the computational resources required to reproduce the reported results.

To the best of our knowledge, this study provides the first {\it detailed numerical exploration of multi-demand, CI-augmented LP converses} for coded caching systems up to $\Ksf=\Nsf=6$ under linear coding assumptions [see GitHub \cite{Brembilla2025CodedCachingLP} for higher-dimensional problems].  
While the current formulation does not yet relax the problem sufficiently to make general $\Nsf{=}\Ksf$ instances computationally tractable, it establishes a reproducible methodology and a scalable formulation that can support future extensions—such as additional CI variables, richer demand symmetries, or alternative relaxation strategies.

%% file: sections/conclusions.tex
\section{Conclusion}
\label{sec:conclusion}

Motivated by the goal of developing a systematic numerical framework to evaluate converse bounds for coded caching, we designed an LP-based approach that leverages known user–file symmetries and entropy equivalences to reduce the dimensionality of the coded caching converse problem. By pruning redundant constraints and merging equivalent variables, the method allows exploration of larger problem instances without significant loss of tightness in the resulting bounds.

The integration of Common Information (CI) variables, extending prior work from the $3$U$3$F case, proved essential for tightening the LP under linear coding assumptions. Together %
with known symmetries, this yields a compact yet expressive LP formulation capable of capturing the key structure of linear coded caching systems.

Using this refined LP formulation, we analyzed problem instances that were previously computationally challenging. These include exact optimality results for the $4$U$4$F and $5$U$5$F problems at $\Msf=1$, and for the $6$U$6$F problem at $\Msf=2$, representing, to the best of our knowledge, the largest problem instances solved to date.
{\bf These results seem to indicate that small, structured demand subsets combined with minimal CI constructions may be sufficient to characterize optimal tradeoffs under linear coding.}

Overall, this work provides a reproducible and scalable methodology for incorporating symmetry and CI structures into LP-based converse formulations.  While generalization to arbitrary $\Nsf=\Ksf$ instances remains computationally demanding, this approach provides a foundation for systematic extensions, such as richer CI models, more sophisticated symmetry reductions, or hybrid relaxations, leading to progressively tighter bounds in high-dimensional coded caching systems.

Once such a scalable numerical framework is established, it can be readily extended to other settings, such as coded caching with heterogeneous cache and file sizes; other shared-link caching models with linear coded placement (e.g., the linear computation broadcast problem); or coded caching with restricted demand types, potentially leading to novel converse bounds under demand-privacy constraints, to name just a few.

\section*{Acknowledgments} 
This work was supported in part by NSF Award 2312229, and used Delta at NCSA through allocation ELE240014 from the ACCESS program~\cite{boerner2023access} supported by
NSF grants 2138259, 2138286, 2138307, 2137603, and 2138296.

%% file: sections/appendix.tex
\section{Analytical Procedure}
\label{sec:AnalyticalProcedure}

The system operates under a NUMA configuration with four domains per socket (\texttt{NPS=4}), improving memory locality and bandwidth utilization during large-scale LP optimization. Gurobi was configured to exploit intra-domain parallelism, minimizing cross-domain memory access and ensuring stable performance across problem instances.  

This Appendix details the analytical process adopted to derive converse bounds prior to presenting the quantitative results in Section~\ref{sec:Results}. Our objective was to identify, for each system configuration $(\Ksf, \Nsf)$, the combination of demand structures and Common Information (CI) variables that yields the tightest linear-program (LP)–based converse bound within the considered memory regime.

\subsection{Problem Instances and Demand Selection}
For each system configuration characterized by \((\Ksf, \Nsf)\) and a given demand type, we begin by constructing the set of distinct demand instances according to the equivalence classes defined in Section~\ref{sec:Methodology}. 
Among these, we identify the subset of demands that yields the largest (worst-case) rate in the corresponding LP relaxation. 
This identification is carried out incrementally: starting from a pair of demands, we successively expand the evaluated subset until no further increase in the resulting rate is observed. 
The resulting subset is thus considered representative of the worst-case for that demand type.

\subsection{Integration of Common Information Variables}
Once the representative demand subset is fixed, the LP is refined through the gradual introduction of Common Information (CI) variables, as defined in Section~\ref{sec:Methodology}. 
At each iteration, a new CI configuration, corresponding to a specific pair or group of variables that share a linear dependence, is incorporated, and the LP is re-solved. 
Due to the rapidly increasing computational complexity, this iterative refinement is continued only until meaningful improvements are observed within practical runtime limits. 
The final rate obtained represents the tightest converse achievable under the tested CI configurations, assuming linear coding for both the placement and delivery phases.

\subsection{Combination Across Demand Types}
The above procedure is repeated for multiple demand types, including the all-file-demanded type demand \(\{1,1,\ldots,1\}\) and composite types such as \(\{2,1,\ldots,1,0\}\) and \(\{3,1,1,\ldots,1,0,0\}\), depending on \((\Ksf,\Nsf)\). 
Each type corresponds to a distinct pattern of user requests and may dominate the converse in different memory regimes. 
To obtain the overall bound, we take the \emph{pointwise maximum} of the rates obtained from all evaluated demand types. 
This ensures that the final tradeoff curve is a valid outer bound, consistent with the methodology of~\cite{tian2018symmetry}.

\subsection{Comparison with Achievability and Converse Benchmarks}
All figures in Section~\ref{sec:Results} report the gap between our refined converse and the known achievability results. 
The achievable tradeoffs are derived from the schemes in Theorems~\ref{thm:performanceYMA} and~\ref{thm:performanceGV}, while the baseline converses follow the results of Theorems~\ref{thm:converseyutheorem2} and ~\ref{thm:converseyutheorem4}. 
Following the convention in~\cite{tian2018symmetry}, the optimal (worst-case) rate for each configuration corresponds to the maximum value among all evaluated demand subsets. 
Accordingly, our figures show the difference between the achievable and converse rates at each normalized cache size~\(\Msf\), where smaller gaps indicate tighter bounds.
Due to the exponential growth of the LP with respect to the number of users, files, and incorporated Common Information variables, the computation was terminated once additional iterations no longer produced significant improvements within a practical runtime.
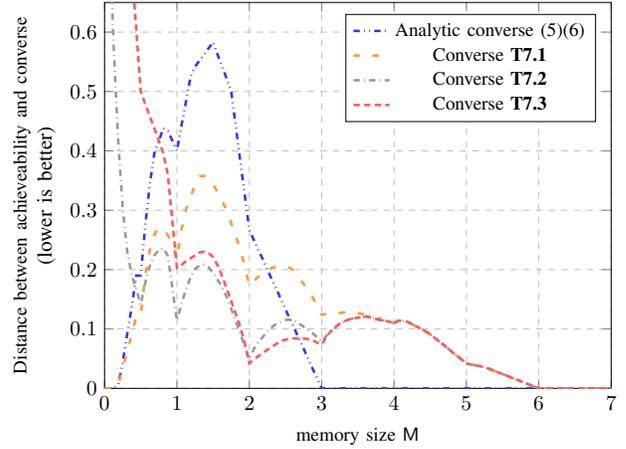
\begin{figure}%[t]
  \centering
  \scalebox{0.8}{
  \begin{tikzpicture}
        \begin{axis}[
            xmin=-0.0, xmax=7.0,
            ymin=-0.0, ymax=0.65,
            legend entries={ 
                \small Analytic converse~\eqref{eq:converseyutheorem2}\eqref{eq:converseyutheorem4},
                \small Converse~\ref{p:t7t1},
                \small Converse~\ref{p:t7t2},
                \small Converse~\ref{p:t7t3},
                \small Converse~\ref{p:t7t4}
            },
            xtick={0,1,...,7},
            ytick={0,0.1,...,0.6},
            y tick label style = {
                /pgf/number format/fixed,
                /pgf/number format/precision=2
            },
            height=8cm,
            width=10cm,
            % compat=1.9,
            grid=major,
            grid style=dashed,
            legend pos=north east,
            ylabel near ticks,
            xlabel={\small memory size $\Msf$},
            ylabel style={align=center},
            ylabel={\small Distance between achieveability and converse \\ (lower is better)},
            scaled ticks=false
        ]
        \addplot[color=blue!80, dash dot dot, very thick] table [x=x, y=y1] {data/7u7f.data};
        \addplot[color=orange!80, loosely dashed, very thick] table [x=x, y=y2] {data/7u7f.data};
        \addplot[color=gray!80, dashdotted, very thick] table [x=x, y=y3] {data/7u7f.data};
        \addplot[color=red!65, densely dashed, very thick] table [x=x, y=y4] {data/7u7f.data};
        \end{axis}
    \end{tikzpicture}
    }
  \caption{7U7F case: achievability  from~\eqref{eq:performanceYMA}-\eqref{eq:performanceGV}. % Tradeoffs~\ref{p:t6t3} and~\ref{p:t6t4} prove that YMA scheme is optimal at $\Msf=2$, and appear to get close to optimality at $\Msf=1$.
  }
  \label{fig:7u7f}
\end{figure}

\begin{figure}
    \centering
  \scalebox{0.8}{
  \begin{tikzpicture}
        \begin{axis}[
            xmin=-0.0, xmax=8.0,
            ymin=-0.0, ymax=0.75,
            legend entries={ 
                \small Analytic converse~\eqref{eq:converseyutheorem2}\eqref{eq:converseyutheorem4},
                \small Converse~\ref{p:t8t1},
                \small Converse~\ref{p:t8t2},
                \small Converse~\ref{p:t8t3},
                \small Converse~\ref{p:t8t4}
            },
            xtick={0,1,...,8},
            ytick={0,0.1,...,0.7},
            y tick label style = {
                /pgf/number format/fixed,
                /pgf/number format/precision=2
            },
            height=8cm,
            width=10cm,
            % compat=1.9,
            grid=major,
            grid style=dashed,
            legend pos=north east,
            ylabel near ticks,
            xlabel={\small memory size $\Msf$},
            ylabel style={align=center},
            ylabel={\small Distance between achieveability and converse \\ (lower is better)},
            scaled ticks=false
        ]
        \addplot[color=blue!80, dash dot dot, very thick] table [x=x, y=y1] {data/8u8f.data};
        \addplot[color=orange!80, loosely dashed, very thick] table [x=x, y=y2] {data/8u8f.data};
        \addplot[color=gray!80, dashdotted, very thick] table [x=x, y=y3] {data/8u8f.data};
        \addplot[color=red!65, densely dashed, very thick] table [x=x, y=y4] {data/8u8f.data};
        \end{axis}
    \end{tikzpicture}
    }
  \caption{8U8F case: achievability  from~\eqref{eq:performanceYMA}-\eqref{eq:performanceGV}. % Tradeoffs~\ref{p:t6t3} and~\ref{p:t6t4} prove that YMA scheme is optimal at $\Msf=2$, and appear to get close to optimality at $\Msf=1$.
  }
  \label{fig:8u8f}
\end{figure}
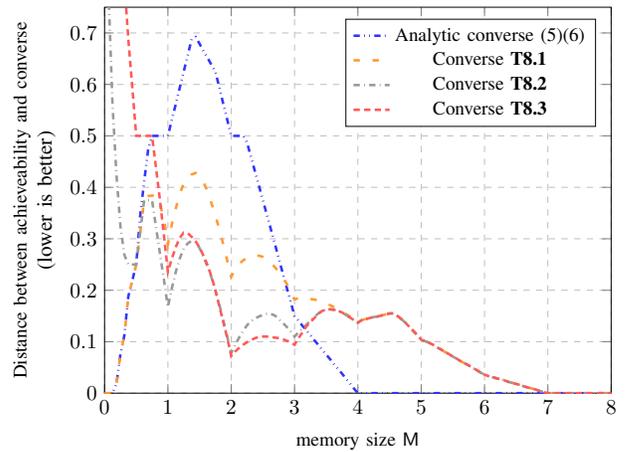

\section{Scaling to Higher-Dimensional Problems}

We further applied the proposed LP-based framework to larger symmetric configurations with $\Ksf = \Nsf$, aiming to identify structural patterns consistent with those observed in smaller systems. To the best of our knowledge, this is the first time converse bounds have been computed for instances as large as 7U7F and 8U8F under linear coding assumptions.

For the 7-user, 7-file case, we explored all demand types combining two demands and one CI variable. 
\begin{enumerate}[label=\textbf{T7.\arabic*}]
\item \label{p:t7t1} Demands
\(\{X_{0123456},X_{1234560}\}\) and auxiliary CI  
\(\{W_0,W_1;Z_0,X_{0123456}\}\).
\item \label{p:t7t2} Demands
\(\{X_{0012345},X_{0123450}\}\) and auxiliary CI \(\{W_0,W_1; Z_0,X_{0012345}\}\).
\item \label{p:t7t3} Demands 
\(\{X_{0001234},X_{0012340}\}\) and auxiliary CI 
\(\{W_0,W_1; Z0,X_{0001234}\}\).
\end{enumerate}

As illustrated in Fig.~\ref{fig:7u7f}, Tradeoff~\ref{p:t7t1} (all-file-demanded type demand) dominates at very low memory values, while Tradeoff~\ref{p:t7t2} prevails in the low-to-mid regime, and Tradeoff~\ref{p:t7t3} achieves the tightest bound around the mid-range.

We extended this analysis to the 8-user, 8-file configuration using two-demand instances. 
\begin{enumerate}[label=\textbf{T8.\arabic*}]
\item \label{p:t8t1} Demands
\(\{X_{01234567},X_{12345670}\}\).
\item \label{p:t8t2} Demands
\(\{X_{00123456},X_{01234560}\}\).
\item \label{p:t8t3} Demands 
\(\{X_{00012345},X_{00123450}\}\).
\end{enumerate}
As shown in Fig.~\ref{fig:8u8f}, the same qualitative pattern emerges: all-file-demanded type demand are most effective at small $\Msf$, and progressively more aggregated request patterns yield tighter bounds as $\Msf$ increases.

Both cases exhibit pronounced dips in the gap at integer memory points, suggesting that the achievable and converse curves may coincide at these points. This indicates that, with enhanced computational resources for solving larger LPs, it may be possible to confirm exact optimality of the YMA scheme at these integer regimes. Conversely, closing the residual gaps at non-integer points would likely require further refinement of both the achievable scheme and the converse formulation.